\documentclass[twocolumn,preprintnumbers,floatfix,amsmath,a4paper,amssymb,nofootinbib, showpacs]{revtex4}

\usepackage{amsmath}
\usepackage{graphicx}
\usepackage{dcolumn}
\usepackage{bm}
\usepackage{amssymb}
\usepackage{epstopdf}
\usepackage{color}

\usepackage
            {hyperref}

\begin{document}

\title{How driving rates determine the statistics of driven non-equilibrium systems with stationary distributions}{}

\author{Bernat Corominas-Murtra$^{1,2}$, Rudolf Hanel$^{1,2}$, Leonardo Zavojanni$^1$ and Stefan Thurner$^{1,2,3,4}$
}
\email{stefan.thurner@meduniwien.ac.at} 

\affiliation{
$^1$ Section for the Science of Complex Systems; CeMSIIS; Medical University of Vienna; 
Spitalgasse 23; A-1090; Vienna, Austria\\
$^2$ Complexity Science Hub Vienna, Josefst\"adterstrasse 39, 1080 Vienna, Austria\\
$^3$ Santa Fe Institute; 1399 Hyde Park Road; Santa Fe; NM 87501; USA\\
$^4$ IIASA, Schlossplatz 1, 2361 Laxenburg, Austria
} 


\begin{abstract}
Sample space reducing (SSR) processes offer a simple analytical way to understand of the origin and ubiquity of power-laws in 
many path-dependent complex systems. SRR processes show a wide range of applications that range from fragmentation processes, 
language formation to cascading processes.
Here we argue that they also offer a natural  framework to understand stationary distributions 
of generic driven non-equilibrium systems that are composed of a driving- and a relaxing process. 
We show that the statistics of driven non-equilibrium systems can be derived from the understanding of the nature of the underlying driving process.
For constant driving rates exact power-laws emerge with exponents that are related to the driving rate. 
If driving rates become state-dependent, or if they vary across the life-span of the process, 
the functional form of the state-dependence determines the statistics.
Constant driving rates lead to exact power-laws, a linear state-dependence function 
yields exponential or Gamma distributions, a quadratic function gives the normal distribution. 
Logarithmic and power-law state dependence leads to log-normal and stretched exponential distribution functions, respectively.
Also Weibull, Gompertz  and Tsallis-Pareto distributions arise naturally from simple state-dependent driving rates. 
We discuss a simple physical example of consecutive elastic collisions that exactly represents a SSR process.
\end{abstract}


\maketitle

\section{Introduction}

\begin{figure}[t]
\includegraphics[width= 8.5cm]{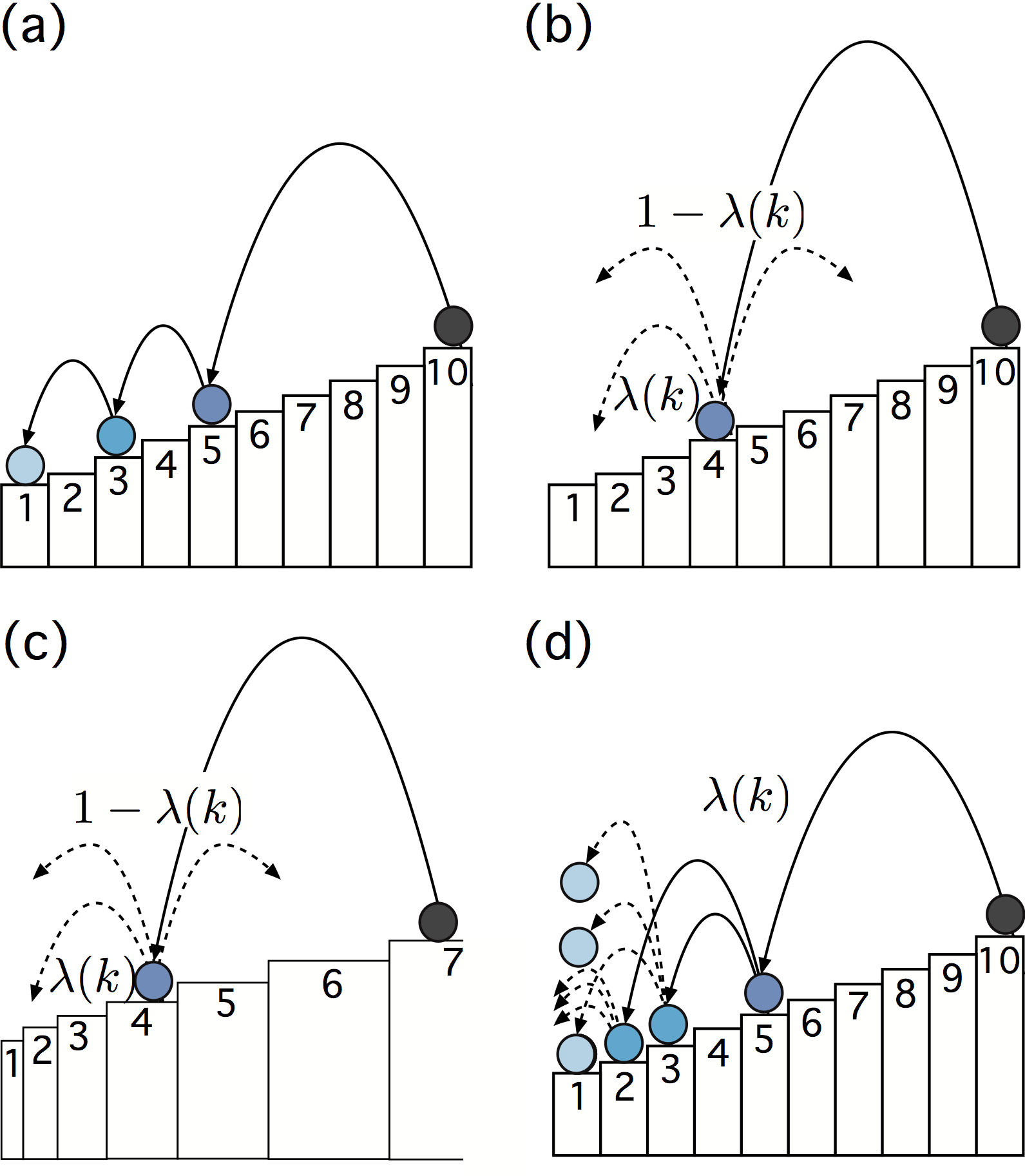}
\caption{
(a) Slowly driven SSR process: relaxation part: a ball bounces downwards a staircase with $N=10$ stairs (states). 
At each timestep the ball randomly choses one of the stairs below its current position. 
In this picture the prior probability of each stair is considered uniform, $q_i=1/N$. 
Driving part: once the ball reaches the lowest step, it is restarted (placed at the highest step $N$, from which it immediately jumps a random step downward -- effectively restarting places it at {\em any} of the $N$ states.) 
The result is Zipf's law in state visits, $p(k)\sim k^{-1}$.  
(b) SSR process with driving: at each step, with probability $1-\lambda$ (driving rate) the ball is restarted, which results in exact 
power-laws,  $p(k)\sim k^{-\lambda}$. 
In the more general setting studied in this paper, driving rates may vary from state to state. 
In the figure the state is $k=4$, and the local driving rate is $1-\lambda(k)$.
(c) One can  assign weights --or prior probabilities $q_i$--  to each state $i$.  
These are represented by different widths of the steps. 
For slow driving, many choices of prior probabilities the histogram of visits to each state shows a perfect Zipf's law, i.e., $p(k)\propto k^{-1}$. 
(d) Whenever $\lambda>1$ we adopt the ``cascading picture'', where, whenever a ball hits a state $i$ it multiplies and creates $\lambda(i)-1$ new balls, that start their downward moves independently. 
For constant $\lambda(i)=\alpha$ we get exact power-law distributions, $p(k)\propto k^{-\alpha}$, with $0\leq \alpha < \infty$ \cite{bernat2}. 
 }
\label{fig:SSESSR}
\end{figure}

Many dissipative systems, driven non-equilibrium processes in particular, can be understood as a combination of  
driving and relaxation processes. 
The relaxation process is characterized by the dynamics that occurs when the system is not driven. 
It describes how the system progresses from ``high'' states (for example energy) towards ``low'' states. 
Without a driving process, the system reaches a stationary ``sink'' or attractor state from which it can no longer escape 
without a driving process. 
The driving process brings the system from low states to high states. 

Typically, relaxation processes  are sample space reducing (SSR) processes, 
meaning that as the system relaxes from higher to lower states, 
the number of possible accessible states reduces over time. 
In other words, when the system is in a high state, there are many lower lying states it can reach. 
When the system is in a low state, it can only reach those few states that are even lower. 
In this sense, the sample space of the relaxing process reduces as the process unfolds. 
When the process is lifted from lower  to higher states by a driving process, the sample space typically increases.

Recently, it was shown that SSR processes exhibit a non-trivial statistical behavior \cite{bernat1}
that allows us to understand the origin and ubiquity of power-laws in many dynamical, 
path-dependent phenomena. 
Examples for SSR processes range from language formation and fragmentation processes  \cite{bernat1,stefan1} 
to diffusion- and search processes on networks \cite{bernat3} to cascading processes \cite{bernat2}. 
SSR processes offer an alternative route to understand power-laws; they complement the classic ways of  
criticality \cite{stanley}, 
self-organized criticality \cite{bak,jensen}, 
multiplicative processes with constraints \cite{gabaix,saichev,malevergne}, 
and preferential processes \cite{yule,simon,barabasi}.

In their simplest form, SSR processes can be depicted as a combination of a relaxing process with a simple driving process. 
For the relaxation process, imagine a ball bounces down a staircase, like the one shown in Fig. \ref{fig:SSESSR}a. 
Each state $i$ of the system corresponds to one particular stair. 
The ball is initially ($t=0$) placed at the topmost stair (highest state $N+1=10$). 
In the next timestep it can jump downward randomly to {\em any} of the $N$ lower stairs, $i=1,2, \cdots N$. 
The probability to hit a particular step $i$ is $q_i= 1/N$. 
Assume that at time $t=1$ the ball landed at  step $k$. Since it can only jump to stairs $k'$ that are 
below $k$, the probability to jump to any stair $k'<k$ in the next timestep is $1/(k-1)$. 
The process continues until eventually stair $1$ is reached; then it halts. 
At this point, the driving process sets in and the process is restarted by placing the ball at state $N+1$ and running a new downward 
relaxation sequence. 
The process can be seen as a generic relaxation process, with a very low driving rate 
that is much slower than the relaxation process.
In this case the frequency of visits to each state $k$ follows an exact power-law 
$p(k)\propto k^{-\alpha}$, with the exponent $\alpha=1$, i.e., Zipf's law \cite{bernat1}.
The existence of Zipf's-law is extremely robust and does not depend on the details of the system.  
It appears as a robust attractor, which emerges for a large variety of prior distributions $q_i$ \cite{bernat3}.  
This means that for these non-uniform prior probabilities $q_i$ 
(which can be interpreted as the width of a stair $i$ in the SSR process, see Fig \ref{fig:SSESSR}c), 
the visiting statistics follows Zipf's law. 
The fact that the power-law is an attractor distribution might explain its ubiquity, in a similar way as the central limit theorem explains the ubiquity of the normal distribution in situations with non-varying sample spaces. 

The power-law exponent can be controlled if the driving rate is increased. 
If the process is restarted\footnote{Think of the restarting process as a process, where the ball is brought to a state $N+1$, 
from which it immediately jumps to any other state, $1,2,\cdots,N$.} 
with probability $1-\lambda$ from any of its current states, 
the distribution function becomes $p(k)\propto k^{-\lambda}$, \cite{bernat1}. 
We call $r=1-\lambda$ the driving rate. 
Intuitively this means if we interrupt the relaxation process with a restarting (driving) event, 
(that brings the system to its highest state $N$) the exponent of the corresponding 
distribution function of states is $1-r=\lambda$. 
This situation is shown in Fig. \ref{fig:SSESSR}b. 
The case of $\lambda=1$ represents the slow driving rate limit mentioned before, 
where the process reaches its lowest state before the restart.  
$\lambda=0$ represents a pure Bernoulli process, since we restart after every step (random walk on the states $1,2,\cdots,N-1$). 

In this picture, $\lambda$ can take any value from $1$ (slow driving) to $0$ (driving at every step). 
Mathematically there is no need to limit the range of $\lambda $ at 1, even though  
for $\lambda >1$ the intuitive picture of the driving rate has to be adjusted. 
However, one can easily interpret a negative driving rate $r$ with a cascading SSR processes 
that is shown in Fig. \ref{fig:SSESSR}d \cite{bernat2}. 
Here $\lambda$ is interpreted as a multiplier that -- whenever a ball reaches a state $i$ -- creates $\lambda-1$ new balls that all sit at state $i$. 
In the next timestep all of these $\lambda$ balls will now enter the relaxation dynamics (bouncing downward in the described fashion), 
creating a cascade of balls during their downward trajectories, Fig. \ref{fig:SSESSR}d. 
We will use the term  ``cascading'' picture instead of the driving rate picture whenever $\lambda>1$.
A negative driving rate means that new random walkers are added at every step with a rate $\lambda$.
For details see \cite{bernat2}. 
These cascading processes show exact power-law distributions $p(k)\propto k^{-\lambda}$ for $0\leq\lambda <\infty$\footnote{For the case of non-uniform priors the situation becomes slightly more involved; the visiting distribution becomes, 
$p(k) \sim q_k/( \sum_{j=1}^k q_j)^{\lambda}$, see \cite{bernat3}.}.

In summary, if the driving rate is zero, meaning that the system reaches a sink state before it is 
lifted to higher states by the driving event, Zipf's law emerges as a robust attractor for the state visit distributions. 
For larger driving rates we obtain exact power-laws, where the exponent is $1-r=\lambda$. 
This also holds for ``negative driving rates'', where $\lambda$ corresponds to the production rate 
of new elements that follow the SSR dynamics. 

Here we will argue that any driven system, for which the relaxing component is sample space reducing,  
the details of the driving component of the system determine the statistics of the state visiting frequencies of the driven system. 
We discuss the case where the driving rate depends locally on the state of the system:  
$1-\lambda(k)$ becomes the driving rate (restarting probability) of the process when at state $k$. 
We show that with particularly simple choices of a state-dependent driving rate $1-\lambda(k)$, 
practically all classical visiting distributions $p(k)$ can be reached, including the exponential, 
normal, Zipf, exact power-law, log-normal, Gamma, Weibull, Gompertz, Tsallis, and power-law with exponential cut-off distributions. 
This view offers a simple generic route to understand stationary distributions of driven non-equilibrium systems 
as a consequence of local or temporal variations of driving rates within a system.
In other words, if the details of a driving process are understood, stationary distributions of driven systems can be predicted.  
We discuss the relation of these results with a recently proposed way to understand various 
stationary distributions on the basis of random growth models \cite{biro}. 

\section{Sample space reducing relaxing processes with state-dependent driving}

Assume a stochastic sample space reducing process over $N$ states with a prior distribution given by any choice of $q_i>0$, 
with $\sum_{i=1}^N q_i =1$. 
If the driving rate is explicitly state-dependent, $1-\lambda(k)$ denotes the probability that the process is restarted  (driven)
whenever it is in state $k$. Let us assume that for all states, $0<\lambda(k)<1$.
The transition probabilities from state $k$ to state $i$ read, 
\begin{eqnarray}
	p_{\rm SSR}(i|k)=\left\{\begin{array}{ll}
	\lambda(k) \frac{ q_i}{g(k-1)}+(1-\lambda(k))q_i\;\;\; {\rm if} \;\;\;i<k \\
	(1-\lambda(k))q_i\;{\rm otherwise}\quad,
\end{array}
\right.
\label{eq:SSRNoiseDep}
\end{eqnarray}
where $g(k)$ is the cumulative distribution of $q_i$, $g(k)=\sum_{i\leq k}q_i$. 
After many restarting events of the relaxing process, one can safely assume the existence of a stationary distribution 
$p_{\lambda,q}$, that depends on the priors and the driving rate. It can be explicitly computed by observing that,
\begin{equation}
	\frac{ p_{\lambda,q}(i+1) }{q_{i+1}}\left(1+\lambda(i+1)\frac{q_{i+1}}{g(i)}\right)=\frac{p_{\lambda,q}(i)}{q_i} \quad .
\end{equation}
We obtain, $p_{\lambda,q}(i)=\frac{q_i}{Z_{\lambda,q}}\prod_{1<j\leq i}\left(1+\lambda(j)\frac{q_j}{g(j-1)}\right)^{-1}$, 
where $Z_{\lambda,q}$ is the normalisation constant. This equation can be well approximated by,
\begin{equation}
	p_{\lambda,q}(i)=\frac{q(i)}{Z_{\lambda,q}}e^{-\sum_{j\leq i}\lambda(j)\frac{q(j)}{g(j-1)}}\quad.
\label{dist}
\end{equation}
If there is a continuum of states, the continuum version of Eq. (\ref{dist}) is, 
\begin{equation}
	p_{\lambda,q}(x)=\frac{q(x)}{Z_{\lambda,q}}e^{-\int_1^x\lambda(y)\frac{q(y)}{g(y)}dy}\quad.
\label{eq:generalSSR}
\end{equation}
Equations (\ref{dist}) and (\ref{eq:generalSSR}) also hold for cascading SSR processes, 
for which $\lambda(k)>1$ plays the role of a state-dependent multiplication rate. 
The case $\lambda(x)=1$ for all $x$, we call a slowly driven SSR process.
Note, that the framework also holds for processes, where not all transitions from all higher to all lower states 
are allowed, but where some are forbidden. 
For constant $\lambda$, this case corresponds to diffusion processes on directed acyclic graphs (targeted diffusion) \cite{bernat3}. 
For state- (node-) dependent $\lambda(x)\leq 1$ on networks, the results derived above hold in the limit of large graphs, 
where $q$ corresponds to the degree sequence. 
The case $\lambda(x)>1$, corresponds to a node-specific multiplication (creation) rate of random walkers, whenever a node is visited.

\subsection{Particular solutions for state-dependent driving}
For simplicity, and with no loss of generality, in the following we consider the continuous case, 
for which we assume that the SSR process is defined on the continuous interval $x \in [1,N]$. 
\begin{figure}
\begin{center}
\includegraphics[width= 8.5cm]{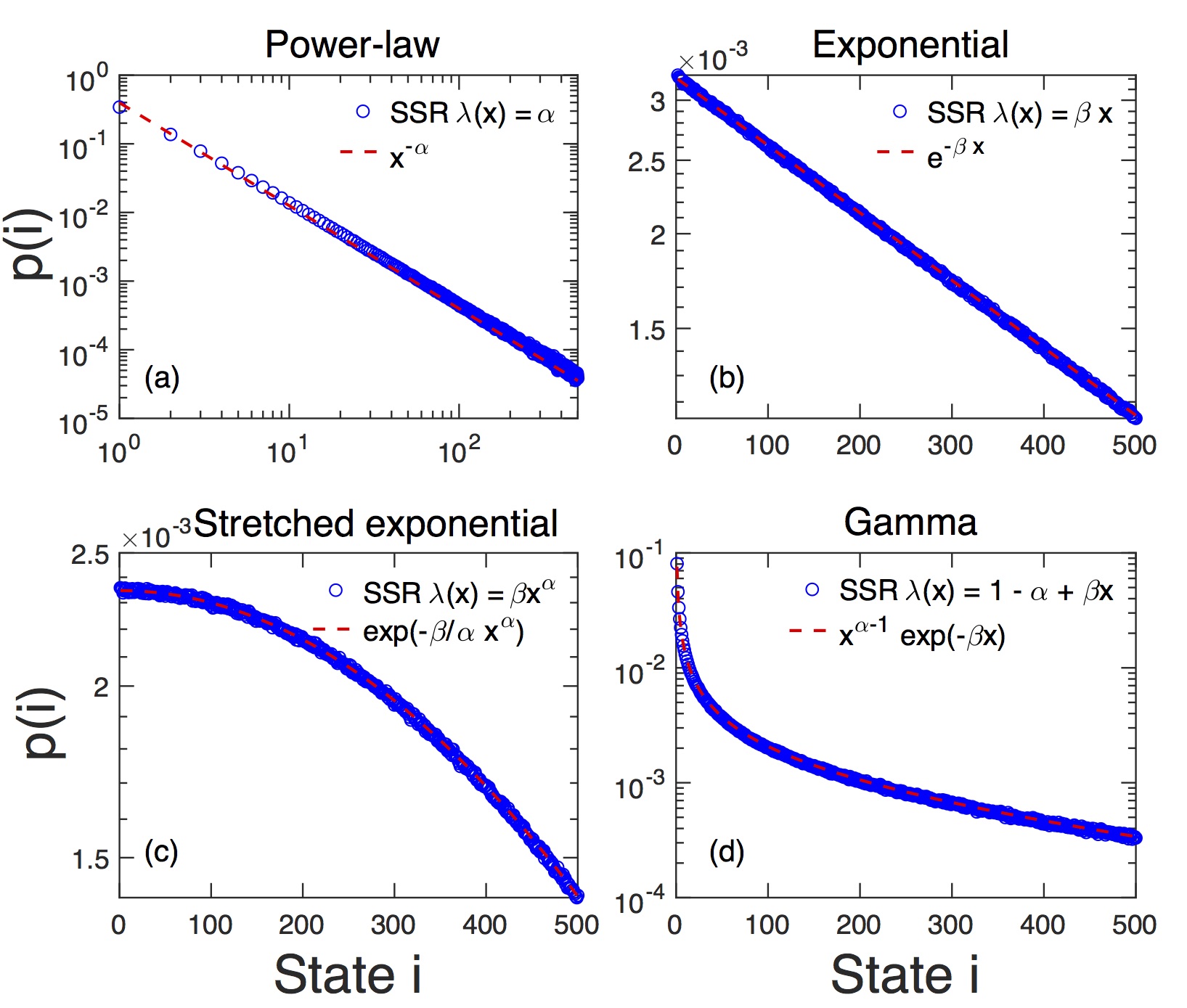}
\caption{
Several classic probability distributions obtained from numerical realisations (circles) of SSR processes over $N=500$ states, 
choosing particular state-dependent driving rates $\lambda(x)$ functions and uniform prior distribution $q$. 
Dashed lines represent the prediction from Eq. (\ref{eq:lambdaSSR}). 
Results are averages over $50$ times $1000$ restarts of the process. Errorbars are generally less than symbol size. 
(a) For constant $\lambda(x)= \alpha$ we obtain exact power-law distributions $p(x) \propto x^{-\alpha}$ ($\alpha = 1.5$). 
In this case, since $\alpha>1$, we have a cascading SSR.
(b) $\lambda(x)= \beta x$ leads to an exponential distribution $p(x) \propto e^{-\beta x}$  ($\beta = 0.00205$).
(c) $\lambda(x)= \beta x^{\alpha} $ leads to a stretched exponential $p(x) \propto e^{-\frac{\beta}{\alpha} x^{\alpha}}$ ($\alpha = 2$, $\beta=4.1{\rm E}-06$). 
Note that $\alpha=2$ corresponds to a normal distribution.
(d) $\lambda = 1-\alpha +\beta x$ yields a Gamma distribution $p(x) \propto x^{\alpha -1} e^{-\beta x}$  ($\alpha = 0.25$, $\beta = 0.0015$).
 }
\label{fig:Numerics}
\end{center}
\end{figure}
We first discuss the case of uniform prior distributions, $q(x)=1/N$. 
To see the relation between the stationary distribution of a process and its driving function $\lambda(x)$, 
we differentiate Eq. (\ref{eq:generalSSR}) and get
\begin{equation}
\lambda(x)=-x\frac{d}{dx}\log p_{\lambda,q}(x) \quad.
\label{eq:lambdaSSR}
\end{equation}
Now we use Eq. (\ref{eq:lambdaSSR}) to compute $\lambda(x)$ for any reasonable distribution function. We present just a few examples, which we  summarize in Table \ref{tab}. Numerical analysis for several driven processes with specific state-dependent noise functions are shown in Fig. \ref{fig:Numerics}, showing perfect agreement with the theoretical predictions.
\\

 \noindent
{\bf Power-laws.} 
\begin{equation}
	p(x)\propto x^{-\alpha}\quad ,
\end{equation}
is obtained with  
$\lambda(x)= -x\frac{d}{dx}\left[-\alpha \log x\right]=\alpha$. 
The fact that a state-independent driving leads to exact power-laws was found in \cite{bernat1}. 
Distributions for state-independent driving are compared with simulations in Fig. \ref{fig:Numerics}a.
\\

 \noindent
{\bf Exponential distribution.}  
\begin{equation}
	p(x)\propto \exp(-\beta x)\qquad  {\rm with} \qquad \beta >0 \quad , 
\end{equation}
is obtained with   
$\lambda(x)= -x\frac{d}{dx}\left[-\beta x\right]=\beta x $. 
Note that for $\lambda<1$,  $\beta\leq 1/N$, in the ``cascading picture'' there is no such upper limit. 
Results from simulated SSR processes with this state-dependent noise are shown in Fig. \ref{fig:Numerics}b.
 \\
 
 \noindent
{\bf Stretched exponential and normal distribution.}  
\begin{equation}
	p(x)\propto \exp\left(-\frac{\beta}{\alpha} x^\alpha\right)
	{\rm with} \quad \alpha >0, \quad \beta >0  \quad ,
\end{equation}
 is obtained with   
$\lambda(x)= -x\frac{d}{dx}\left[-\frac{\beta}{\alpha} x^\alpha\right]= \beta x^\alpha$. 
$\alpha=2$ corresponds to the normal distribution.
Again, $\lambda<1$ implies $\beta\leq N^{-\alpha}$, while  
in the ``cascading picture''  no such limitation exits,  see Fig. \ref{fig:Numerics}c.
\\

 \noindent
{\bf Gamma distribution.}
\begin{equation}
	p(x)\propto x^{\alpha-1}\exp(-\beta x) \quad  {\rm with} \quad \alpha>0, \quad \beta >0 \quad ,
\end{equation}
is obtained with   
$\lambda(x)= -x\frac{d}{dx}\left[(\alpha-1)\log x-\beta x\right] =1-\alpha+\beta x$. 
Obviously, $\alpha-1\leq \beta$ is required. See Fig. \ref{fig:Numerics}d.
\\

\noindent
{\bf Log-normal distribution.} 
\begin{equation}
	p(x)\propto \frac{1}{x}e^{-\frac{(\log x-\beta)^2}{2\sigma^2}}\quad ,
\end{equation}
is obtained with  
$\lambda(x)= -x\frac{d}{dx}\left[-\log x- \frac{(\log x-\beta)^2}{2\sigma^2}\right]=1+\frac{\log x}{\sigma^2}-\frac{\beta}{\sigma^2}$. 
For $\lambda<1$ we require $\log N\leq \beta \leq \sigma^2$.
\\

 \noindent
{\bf Power-law with exponential cut-off.}
\begin{equation}
	p(x)\propto x^{-\alpha}\exp(-\beta x) \quad  {\rm with} \quad \alpha>0, \quad \beta >0 \quad ,
\end{equation}
is obtained with  
$\lambda(x)= -x\frac{d}{dx}\left[-\alpha\log x-\beta x\right]=\alpha+\beta x$. Here
$\lambda<1$ implies the restrictions  $\beta \leq (1-\alpha)/N$ and $\alpha\leq 1$.
\\
 
 \noindent
{\bf Tsallis-Pareto or $q$-exponential distribution.}
\begin{equation}
	p(x)\propto (1-(1-Q)\beta x)^{\frac1{1-Q}} \quad  {\rm with} \quad \beta >0 \quad ,
\end{equation}
is obtained with  \\
$\lambda(x)= -x\frac{d}{dx}\left[\frac1{1-Q}\log(1-(1-Q)\beta x)\right] 
= \frac{\beta x}{1-\beta x(1-Q)}$. 
Note that for for $q$-exponentials with $Q<1$ we require $\beta x <1/(1-Q)$,
while for $Q>1$ no such restriction exists. For $\lambda<1$ we also require,  
$(2-Q)\beta\leq 1/N$, which for $Q>2$ is always satisfied. 
\\
   
 \noindent
{\bf Weibull distribution.} 
\begin{equation}
	p(x)\propto x^{\alpha-1}e^{-\left(\beta x\right)^\alpha} \quad  {\rm with} \quad \alpha > 0, 
	\quad \beta >0 \quad , 
\end{equation}
is obtained with  \\
$\lambda(x)= -x\frac{d}{dx}\left[(\alpha-1)\log x-\left(\beta x\right)^\alpha\right]=1-\alpha+\alpha
\left(\beta x\right)^\alpha$. \\
Note that the standard parametrization of the Weibull distribution uses the parameter $\nu=1/\beta$
instead of $\beta$. To ensure positivity of $\lambda$ this implies that
$ \frac{\alpha-1}{\alpha}\leq \beta^{\alpha}$,  which is always satisfied for $\alpha<1$. 
For $\alpha>1$ we need $\beta\leq \left(1-\frac1\alpha\right)^{\frac1\alpha}$, for $\alpha<1$ we require $\beta\leq 1/N$. 
\\
 
 \noindent
{\bf Gompertz distribution.} 
\begin{equation}
	p(x)\propto \exp\left(\beta x-\eta e^{\beta x}\right) \quad  {\rm with} \quad \beta > 0, \quad \eta >0 \quad , 
\end{equation}
is obtained with 
$\lambda(x)= -x\frac{d}{dx}\left[\beta x-\eta e^{\beta x}\right]=(\eta e^{\beta x}-1) \beta x$. 
The restriction $e^{-\beta}\leq \eta$ applies, for the noise picture we further require   
$\eta\leq \left(\frac1\beta+1\right)e^{-\beta}$. 

 \begin{table}[ht!]
\caption{Relations between state-dependent driving functions $\lambda(x)$ and distribution functions $p_{\lambda,q}(k)$ for driven SSR processes.}
\begin{tabular}{l l l } 
\hline
distribution &  $\lambda(x)$ &  $p_{\lambda,q}(x)$  \\ 
 \hline
 Power-law 			& $\alpha$ 							& $ x^{-\alpha}$ \\ 
 Exponential			& $\beta x$ 		 					& $ e^{-\beta x}$  \\
 Power-law with cut-off 	& $\alpha+\beta x$ 						& $ x^{-\alpha}e^{-\beta x}$ \\
 Gamma				& $1-\alpha+\beta x$ 					& $ x^{\alpha-1}e^{-\beta x}$  \\
 Log-normal 			& $1-\frac{\beta}{\sigma^2}+\frac{\log x}{\sigma^2}$ & $\frac{1}{x}e^{-\frac{(\log x-\beta)^2}{2\sigma^2}}$\\
 Normal ($\alpha=2$) 	& $\beta x^2$ 							& $ e^{-\frac{\beta}{2}x^{2}}$  \\
 Stretched exponential 	&$\beta x^\alpha$ 				& $ e^{-\frac{\beta}{\alpha}x^{\alpha}}$  \\
 Gompertz 			& $(\alpha e^{\beta x}-1)\beta x$ 			& $ e^{\beta x-\alpha e^{\beta x}}$ \\ 
 Weibull 				& $1-\alpha+\alpha\left(\beta x\right)^\alpha$ 	& $ x^{\alpha-1}e^{-\left(\beta x\right)^\alpha}$ \\ 
 Tsallis-Pareto			& $\frac{\beta x}{1-\beta x(1-Q)}$			& $(1-(1-Q)\beta x)^{\frac1{1-Q}}$\\
 \hline
\end{tabular}
\label{tab}
\end{table}

\subsection{Non-uniform prior distributions}

In general, for the case in which both, $\lambda$ and $q$ are functions of the state $x$, 
a unique relation between driving and distribution functions is hard or even impossible to find. 
However, some cases can be explored. 
As an example we show the situation for the specific driving function, $\lambda(x)=\beta g(x)$. 
Using Eq. (\ref{eq:generalSSR}) we have, 
\begin{equation}
	p_{\lambda,q} (x) =\frac{q(x)}{Z_{\lambda,q}}e^{-\beta g(x)}\quad.
\end{equation}
Taking the derivative we get,
\[
\frac{d}{dx}\log p_{\lambda,q}(x)=\frac{d}{dx}\log q(x) -\beta q(x)\quad,
\]
which has the general solution, 
\begin{equation}
q(x)=\frac{\beta ^{-1}p_{\lambda,q}(x)}{c-P_{\lambda,q}(x)}\quad,
\label{qq}
\end{equation}
where $c>1$ is a constant and $P_{\lambda,q}$ is the cumulative distribution associated to $p_{\lambda,q}$. 
Once  $p_{\lambda,q}$ and $\lambda(x)$ (which is equivalent to the cumulative $g(x)$, up to a constant $\beta$)
are specified, $q(x)$ is computed through Eq. (\ref{qq}).

 \section{A physical example}

Repeated elastic collisions of spherical projectiles with targets of identical masses in three dimensional space
are an exact example for a continuous ``staircase process'' with uniform priors\footnote{Except for minor technicalities, 
	SSR processes on continuous sample spaces behave exactly as discrete SSR processes.}.
Assume a simple experiment in which projectiles are fired into a container that consists of 
$D$ layers of targets. A projectile might be an atom and the container is a foil with $D$ layers of target atoms.

Every time the projectile collides elastically with a target (in rest), it transfers some of its kinetic energy to the target.
In sequences of collisions the projectile's kinetic energy reduces after every collision and follows a SSR dynamics. 
The transition probability density to find the projectile with energy $E'$ after a collision, given that it entered the collision with energy $E$, is 
\begin{equation}
	\rho_{\rm SSR}(E'|E)=\frac{\theta(E-E')}{E}\quad,
\end{equation}
where $\theta$ is the Heaviside step-function, see appendix \ref{appA}. 
This reminds us immediately of Eq. (\ref{eq:SSRNoiseDep}), for $\lambda=1$, and $q_i$ uniform. 
Let  $p_c$ denote the collision probability that a projectile while passing through a layer collides with a target, then 
on average a projectile will encounter $p_cD$ collisions on its path through the foil. A SSR process with $r=1-\lambda$, 
where $r$ is what we called the {\em driving rate}, will on average perform $(1-r)/r$ SSR steps before it leaves the foil and 
a new projectile is fired (restart). 
We identify $p_cD\sim (1-r)/r$ and get $r\sim 1/(1+p_cD)$.  
If $p_c$ is a constant, the empirical distribution of projectile kinetic energies
sampled after collisions on their path through the target is described by
\begin{equation}
	p(E)\propto E^{-\frac{p_cD}{1+p_cD}}\quad . 
	\label{eq:zipfcollision}
\end{equation}
This implies that for thick foils ($D\to \infty$) we get Zipf's law with $\lambda\sim 1$. 
Note that we do not take travel times between layers into account. 

In many elastic collision experiments, such as neutron scattering, the collision probability  $p_c=p_c(E)$  
is energy-dependent due to energy-dependent {\em cross-sections}.   
In these cases $\lambda(E)$ is a state (energy)-dependent property of the process. 
As a consequence, the observable distribution functions after multiple collisions will follow our central result in 
Eq. (\ref{eq:generalSSR}).
 
Note that the power-law in Eq. (\ref{eq:zipfcollision}) directly translates to the well-known exponential 
energy profiles in absorbing media. 
The expectation values of the projectile's energies $E_n$ after the $n$'th subsequent collision indeed follow an exponential, 
\begin{equation}
	\langle E_N\rangle=E_0e^{-\beta N}\quad ,
\end{equation}
where $E_0$ is the initial kinetic energy of projectiles, see appendix \ref{appB}.
For uniform priors one gets $\beta=\log 2$, i.e. $\exp(-\beta)=1/2$.
In our thought experiment with a constant collision probability $p_c$, a projectile that has traveled a 
distance $d$ ($0\leq d\leq D$) through the container has undergone $N \sim p_cd$ collisions,   
and the average projectile energy is, $\langle E(d)\rangle=E_0e^{-\beta p_c d}$.
This exponential law also reminds us of the Lambert-Beer law that  
describes the loss of intensity of radiation traveling through an absorbing medium. 
For the Lambert-Beer law we may conversely conclude that the intensity distribution of radiation itself follows a power-law with an exponent depending
on the absorption coefficient and the thickness $D$ of the absorbing medium. 
Moreover, in inhomogeneous media varying absorption coefficients again allow the parameter $\lambda$ of the process to become 
state-dependent.

\section{Discussion}
Driven non-equilibrium systems are often composed of a driving process and a relaxing process. 
The later is characterized by transitions from higher states to lower states, and is often a 
sample space reducing process. 
SSR processes with simple driving processes have been shown to be analytically solvable. 
They exhibit non-Gaussian statistics that is often encountered in driven complex systems. 
In particular SSR processes offer an alternative route to understand the origin of power-laws. 
Here we showed that SSR processes exhibit a much wider range of statistical diversity if the driving process becomes non-trivial. 
Assuming that  driving rates may vary with the current state of the system, 
we demonstrated that practically any distribution function can be naturally associated with state-dependent driving processes. 
The functional form of the driving function can be extremely simple. Constant driving leads to exact power-laws, 
a linear driving functions $\lambda(x)$
gives exponential or Gamma distributions, a quadratic function yields the normal distribution. Also the Weibull and Gompertz 
distributions arise as a consequence from relatively simple driving functions. 
It is well known how noise and drift parameters can be defined in standard stochastic processes to derive specific stationary distributions. 
In this sense, note that Eq. (\ref{eq:generalSSR}) is also the solution of a general family of stochastic differential equations 
\cite{VanKampen:2007}, where the drift and noise terms are defined in terms of $q$, $g$ and $\lambda$ in the following way,
\begin{equation}
dX(t)=-\frac{1}{2}\frac{\lambda(x)}{g(x)}dt+ q(x)^{- \frac12} dW\quad,
\end{equation}
where $dW$ defines a Wiener process. 
The relation between standard stochastic equations and driven SSR process with general state-dependent noise is purely formal. 
The underlying Wiener process is qualitatively different from the microscopic dynamics of a relaxing SSR processes. 
In driven SSR processes there is a straightforward and clear interpretation of all parameters involved.

More interesting than this formal correspondence to stochastic processes, 
is the relation of the state-dependent driven processes with sustained random growth models (SRG) \cite{biro}.
Similar to driven SSR processes, SRG processes also cover a wide range of real-world applications. 
These are processes, where random walkers run through a directed chain of states. 
The transition rate from one state $n$ to the next state $n+1$ is labelled by 
$\mu_n$. At every state there is a probability $\gamma_n$ that the walker leaves the chain and disappears. 
The process is sustained by a constant inflow of walkers to the first state. 
A remarkable feature of this system is that it exhibits stationary distributions, 
\begin{equation}
	p_{\mu,\gamma}(x)=\frac{1}{Z_{\mu,\gamma}} \mu(x)^{-1} e^{-\int_1^x\frac{\gamma(y)}{\mu(y)}dy} \quad,
\label{eq:Biro}
\end{equation}
where $Z_{\mu,\gamma}$ is the normalisation constant. Equation (\ref{eq:Biro}) has a  similar structure to Eq. (\ref{eq:generalSSR}). 
Indeed one can map one-to-one $\lambda(x)$ and $q(x)$  of the driven SSR to the $\mu(x)$ and $\gamma(x)$ of the SRG. 
By specifying specific relations between $\mu$ and $\gamma$, the SRG allows us to derive a large variety of distribution functions. 
In particular, the mapping can be obtained by setting
$\lambda(x)= \gamma(x)\int_1^x\frac{dy}{\mu(y)}$ and $q(x) = \left( \mu(x) \int_1^N \frac{dy}{\mu(y)} \right)^{-1}$. 
The reverse relation is given by $\mu(x)=\frac{Z_{\mu,\gamma}}{Z_{\lambda,q}}\frac{1}{q(x)}$ and $\gamma(x)=\frac{Z_{\lambda,q}}{Z_{\mu,\gamma}}\frac{\lambda(x)}{g(x)}$. For constant driving, the SRG parameters $\mu(x)$ and $\gamma(x)$ are the inverse of $q(x)$ and $g(x)$, respectively. 
Accordingly, the existence of this mapping enables us to relate a sampling process with a collapsing sample space (driven SSR processes) 
with a stochastic process that runs in the opposite direction (SRG) and populates more states as it unfolds.   
We demonstrated that SSR processes occur not only in complex systems but already in simple consecutive elastic collision experiments. 
Collisions in materials with energy-dependent cross sections exactly correspond to our main result of how state-dependent $\lambda$
correspond to observable distribution functions.

$\\$
The authors declare no competing financial interest.

$\\$
We acknowledge support from Austrian Science Foundation, under FWF projects P29032 and P29252. 

$\\$
B C-M, R H, L Z and S T contributed equally to conceive and design the experiments, develop mathematical models, perform numerical experiments and write the paper.

\appendix

\section{Elastic collisions as a SSR processes \label{appA}}

Consider an elastic collision of two particles of mass $m$ (projectile) and $M$ (target) with respective radii $r$ and $R$. 
In the center of mass system the projectile and target have initial velocities $u$ and $v$, respectively. 
Momentum conservation implies $v=-um/M$. 
We assume that both particles move along the $x$ axis of the center of mass coordinate system.
After the collision, both particles move along with the same velocities but with their directions may have changed by an angle 
$\phi$, i.e. $u'=-u (\cos(\phi),\sin(\phi))$ and $v'=-v (\cos(\phi),\sin(\phi))$. 
The velocities of the particles in the laboratory coordinate system, where the target particle with mass $M$ is in rest before the collision, 
one finds $v_0=0$. It follows that $u_0=u-v=u(1+m/M)$. 
Moreover, after the collision, $v_1=v'-v(1,0)=u(\cos(\phi)+1,\sin(\phi))m/M$ and $u_1=u'-v(1,0)=u(m/M-\cos(\phi),-\sin(\phi))$. 
If the projectile moves at speed $u_0$ before the collision, then after the collision it moves at speed
$u_1=u_0(m/M-\cos(\phi),-\sin(\phi))/(1+m/M)$. The kinetic energy of this particle before the collision
is $|u_0|^2=2E_0/m$ and after the collision $|u_1^2|=2E_1/m=(2E_0/m)((m/M)^2-2m/M\cos(\phi)+1)/(1+m/M)^2$. 
It follows that 
\begin{equation}
\begin{array}{lcl}
	E_1(\phi)&=&E_0\frac{\frac{m}{M}+\frac{M}{m}-2\cos(\phi)}{(\frac{M}{m}+2+\frac{m}{M})}\\
	&=& E_0\left(1-\frac{2}{\mu}(\cos(\phi)+1)\right)\quad ,
\end{array}
\end{equation}
with $\mu=\frac{(M+m)^2}{mM}$ and
\begin{equation}
	dE_1(\phi)=E_0\frac{2}{\mu}\sin(\phi)d\phi\quad .
	\label{app:dE}
\end{equation}
Clearly, $E_1$ may take values in the interval $[qE_0,E_0]$,
where $q=1-4/\mu=\left(\frac{M-m}{M+m}\right)^2$.
For $m=M$, $q=0$ and $E_1\in[0,E_0]$.




{\bf Transition probabilities in three dimensions.}
Defining again $R_c=R+r$ one gets that if a projectile hits the target at a distance $x$ off their centers, $0<x<R_c$, is $\rho_x(x)=2x/R^2_c$, i.e. $P_x([x,x+dx])=\rho_x(x)dx=2xdx/R^2_c$. The tangential angle $\alpha$ between the colliding particle is $\cos(\alpha)=x/R_c$; and the reflection angle is $\phi=2\alpha$. As a consequence $x=R_c\cos(\phi/2)$ and
$dx=-R_c\sin(\phi/2)/2\ d\phi$. From $P_\phi([\phi,\phi+d\phi])=\rho_\phi(\phi)d\phi=-\rho_x(x)dx=P_x([x+dx,x])$ we get
\begin{equation}
	\rho_\phi(\phi)=\frac12\sin\left(\phi\right)\,.
\end{equation}

From $\rho_\phi(\phi)d\phi=\rho(E_1(\phi)|E_0)dE_1(\phi)$ and Eq. ({\ref{app:dE}), we get
\begin{equation}
	\rho(E_1|E_0)=\frac{\mu}{4E_0}\quad .
\end{equation}
We see that the transition probabilities in $E$ for elastic collisions and $M=m$ follow exactly the typical SSR dynamics. 

\section{Zipf and exponential distribution functions \label{appB}}

For (continuous) SSR processes, where prior weights of the process are given by a power-law $q(E)\propto E^{\alpha}$ with exponent $\alpha>-1$, the energy expectation values $\langle E_n\rangle$ of the projectile energies $E_n$ after the $n$'th SSR move, decay exponentially. 
To see it we define the cumulative distribution $Q(E)=\int_0^{E} dE' q(E)$ and for simplicity we first show the case for $\alpha=0$, 
\begin{equation}
\begin{array}{lcl}
	\langle E_{n+1}\rangle&=&\int_0^{E_0}dE'E' \rho_{n+1}(E'|E_0)\\
	&=&\int_0^{E_0}dE'E' \int_{E'}^{E_0}dE_{n}\rho_{1}(E'|E_{n})\rho_{n}(E_{n}|E_0)\\
	&=&\int_0^{E_0}dE'E' \int_{E'}^{E_0}dE_{n}\frac{1}{E_n}\rho_{n}(E_{n}|E_0)\\
	({\rm part.\ int.})&=&\frac{1}{2}\int_0^{E_0}dE'E'^2\frac{\rho_{n}(E'|E_0)}{E'}\\
	&=&\frac{1}{2}\int_0^{E_0}dE'E'\rho_{n}(E'|E_0) =\frac{1}{2}\langle E_n\rangle
\end{array}
\end{equation}
For $\alpha>-1$ the computation follows exactly the same logic,
and we get $\langle E_{n+1}\rangle=\frac{\alpha+1}{\alpha+2}\langle E_n\rangle$. 
As a consequence one finds that 
\begin{equation}
	\langle E_{n}\rangle = E_0\left(\frac{\alpha+1}{\alpha+2}\right)^{n}=E_0 e^{-\beta n}\quad,
	\label{app:exp}
\end{equation}
with $\beta=\log(2+\alpha)-\log(1+\alpha)$.
To translate the $n$'th collision into a penetration depth $d=1,2,\cdots,D$, for a constant $p_c$, 
just use $n=p_c d$ in Eq. (\ref{app:exp}).

\end{document}